\documentclass[10pt, conference, twocolumn, letterpaper]{IEEEtran}

\usepackage{booktabs} 

\usepackage{tabularx}
\usepackage[ruled,linesnumbered]{algorithm2e}
\usepackage{booktabs}
\usepackage{epsfig}
\usepackage{amsmath}
\usepackage{subfigure}
\usepackage{graphicx}
\usepackage{cite}
\usepackage{amssymb}
\usepackage{fixmath}
\usepackage{caption}
\usepackage[usenames]{color}
\usepackage{dsfont}
\usepackage[super]{nth}
\usepackage{color,soul}

\usepackage{multirow}
\usepackage{comment}
\usepackage{balance}
\usepackage{paralist}

\usepackage{graphicx}
\usepackage{subcaption}

\usepackage{float}
\usepackage{dblfloatfix}
\usepackage{siunitx}
\usepackage{array}

\newdimen\figrasterwd
\figrasterwd\textwidth

\usepackage{flushend}

\usepackage{makecell} 

\usepackage{arydshln}

\usepackage{fancyhdr}
\IEEEoverridecommandlockouts 

\begin{document}




\title{C-POD: An AWS Cloud Framework for Edge Pod Automation and Remote Wireless Testbed Sharing}

\author{Annoy Dey$^1$, Vineet Sreeram$^2$,  Gokkul Eraivan Arutkani Aiyanathan$^2$, Maxwell McManus$^2$,\\ Yuqing Cui$^1$, Guanying Sun$^2$,  Elizabeth Serena Bentley$^3$, Nicholas Mastronarde$^2$, and Zhangyu Guan$^1$\\
$^1$Dept. CSE, University of Minnesota-Twin Cities, USA; 
$^2$Dept. EE, University at Buffalo, USA; 
$^3$U.S. AFRL\\
Email: 
\{dey00063, cui00236, zguan\}@umn.edu,
elizabeth.bentley.3@us.af.mil,\\
\{vsreeram, gokkuler, memcmanu, gsun4, nmastron\}@buffalo.edu
\thanks{This work was supported in part by the National Science Foundation (NSF) under Grant SWIFT-2229563 and CNS-2450418, and the U.S. Air Force Research Laboratory under Contracts FA8750-21-F-1012, FA8750-20-C-1021 and FA8750-25-1-1000. Distribution A. Approved for public release: Distribution Unlimited: AFRL-2025-4262 on 25 Aug 2025.}
}

\maketitle

\thispagestyle{fancy} 


\begin{abstract}
This paper presents C-POD, a cloud-native framework that automates the deployment and management of edge pods for seamless remote access and sharing of wireless testbeds. C-POD leverages public cloud resources and edge pods to lower the barrier to over-the-air (OTA) experimentation, enabling researchers to share and access distributed testbeds without extensive local infrastructure. A supporting toolkit has been developed for C-POD to enable flexible and scalable experimental workflows, including containerized edge environments, persistent Secure Shell (SSH) tunnels, and stable graphical interfaces. We prototype and deploy C-POD on the Amazon Web Services (AWS) public cloud to demonstrate its key features, including cloud-assisted edge pod deployment automation, elastic computing resource management, and experiment observability, by integrating two wireless testbeds that focus on RF signal generation and 5G(B) communications, respectively. 

\end{abstract}


\section{Introduction} \label{sec:intro}

OTA experiments on testbeds are a key step in bridging the gap between theory and practice in wireless research. These experiments enable researchers to validate models, identify implementation challenges, and refine designs in controlled but realistic environments. In recent years, significant efforts have been made by the wireless research community toward the development of such platforms. Notable examples include the NSF PAWR platforms POWDER \cite{breen2020powder}, COSMOS \cite{rauchaudhari2020cosmos}, AERPAW \cite{marojevic2020aerpaw}, and ARA \cite{zhang22ara}, Colosseum \cite{bonati2021colosseum}, and the NextGen end-to-end network testbed \cite{Chouman2024EndtoEnd}, among others.

Although these large-scale testbeds have significantly advanced wireless experimental research, they typically support only a narrow range of research topics and are confined to fixed, pre-selected deployment environments. In contrast, many grassroots wireless testbeds developed by individual groups have explored a wide range of research topics in heterogeneous environments. However, it should be noted that remotely sharing such testbeds between research groups is complex and costly.
Individual small research groups usually lack the financial and technical resources to build the necessary local computing and storage infrastructure and dedicated web portals \cite{gomez2023survey}. In addition, long-term operation requires sustained hardware maintenance, regular software updates, and persistent user support. To date, the wireless community still lacks a mature and convenient approach for research groups to share their testbeds with collaborators and the broader community.

To bridge this gap, built on our previous work \cite{CloudRAFT, NEXTCOMCOM, UnionLabs, WiNTECH24},  in this paper we further introduce C-POD, a cloud-native framework that automates the deployment and management of edge pods for remote wireless testbed access. Using a combination of clod and edge resources, C-POD reduces the barrier to OTA experimentation by enabling researchers to share and access geographically distributed testbeds without extensive local infrastructure. It is particularly beneficial for small research groups that lack the resources or expertise to develop and maintain dedicated web portals and testbed management infrastructure. With its ready-to-use cloud native toolkit, C-POD makes these testbeds remotely accessible and easily shareable, enhancing their utility and impact. Moreover, C-POD supports dynamic edge scaling, persistent remote sessions, and stable graphical interfaces, making it well suited for long-running and high-variability experiments.

 
The main contributions of this work are as follows.
\begin{itemize}
\item \textit{Cloud-Enabled Edge Pod Automation Framework:} We design a cloud-native, end-to-end framework based on Amazon Web Service (AWS) and Rancher Kubernetes Engine v2 (RKE2) \cite{rke2docs} to provide isolated, resource-capped experiment pods on demand at the testbed edge. 

\item \textit{Elastic Resource Management Toolkit:} We develop a toolkit for real-time auto-scaling of containerized edge services using AWS Lambda, enabling elastic resource allocation based on workload demands during experiments. 

\item \textit{Testbed Integration and Experimental Validation:} We deploy C-POD in the AWS cloud and integrate it with wireless testbeds comprising various software-defined radios (SDRs), including USRP X310/B210 units for EPC and gNB, B210 for User Equipment (UE) and USRP N210s for waveform generation and spectrum sensing. 
\end{itemize}

\begin{figure*}[t]
    \centering
    \includegraphics[width=0.72\textwidth]{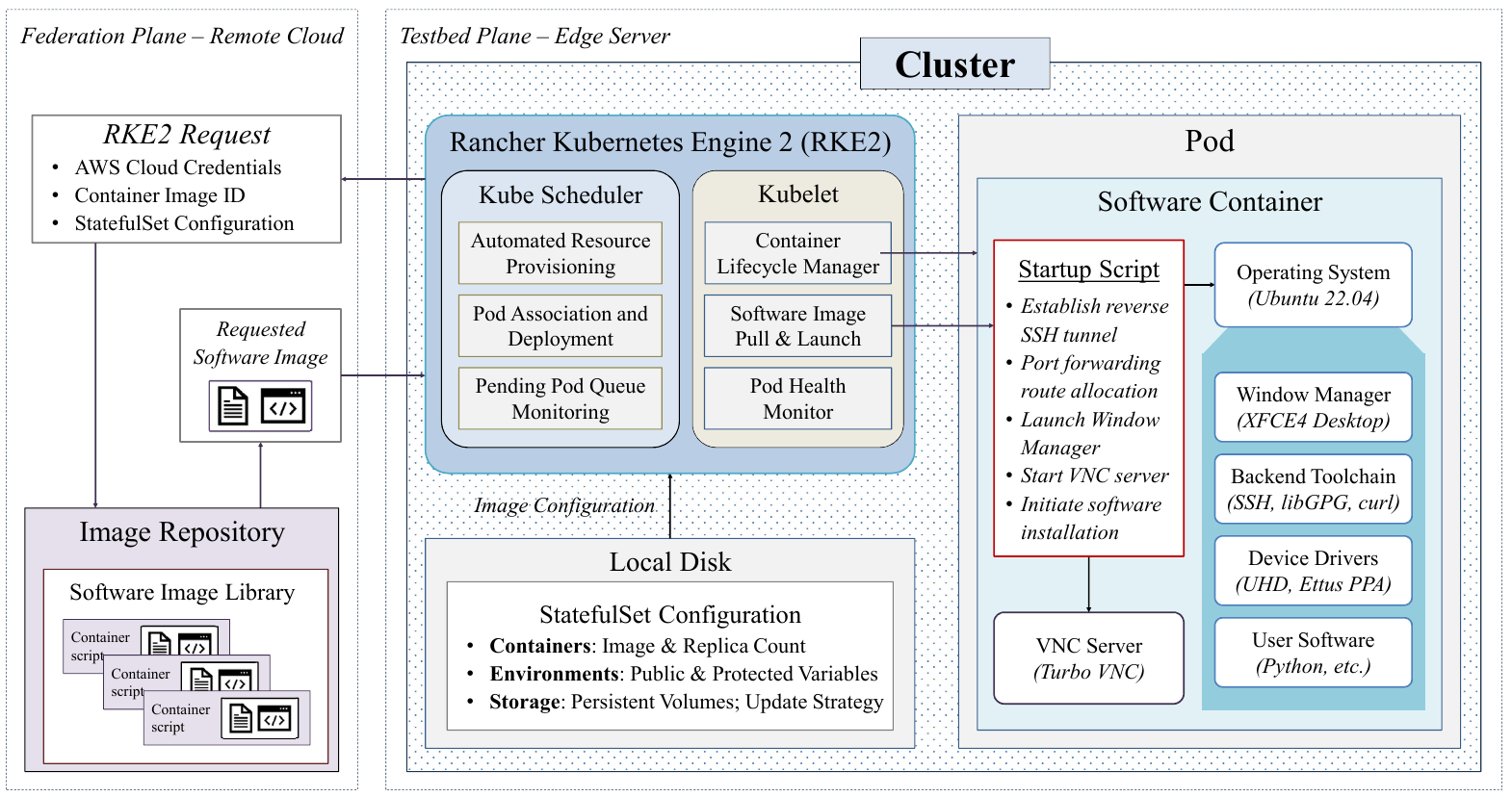}
    \caption{\small Cluster and pod structure in C-POD: Illustration of interactions between RKE2 components, container lifecycle management, and the software environment, including StatefulSet configuration, image pulling, reverse SSH/VNC setup, and automated resource provisioning. \vspace{-4mm}}
    \label{fig:Podarch}
\end{figure*}


 

\section{C-POD Framework and Toolkit Design} \label{sec:Federation}

The C-POD framework, illustrated in Fig.~\ref{fig:Podarch}, consists of two primary components: the \textit{Federation Plane} and the \textit{Testbed Plane}. 
The \textit{Federation Plane} serves as an elastic layer that provides a granular mapping between user requests and edge testbeds. Requests submitted via the web portal are first directed to the Interface Gateway, which uses an API management module to handle authentication, authorization, and endpoint exposure. The \textit{Federation Service Interface Module} then mediates communication between web portal and back-end services, while the \textit{User Traffic Routing Module} ensures that each user is connected to the appropriate service instance.

Upon receiving user requests, a key function of the \textit{Federation Plane} is to initiate the appropriate service automation workflows. This process is managed by the \textit{Function Manager Module}, which interacts with all other components of the plane, such as the storage repository, database engine, and operations manager, through dedicated APIs. For example, 
when a user request involves accessing the \textit{Storage Repository}, the \textit{Function Manager} coordinates file transfers by utilizing temporary storage for fast data movement and community libraries for commonly used code. If the services initiated by the \textit{Function Manager} require container images, the \textit{Image Repository} is invoked to fetch the requested container images.

Finally, the \textit{Operations Manager Module} orchestrates interactions with edge devices within the \textit{Testbed Plane} by deploying the SSM agent on the testbed’s control plane and maintaining continuous communication with it. This module also coordinates with the elastic VM to launch the noVNC client for browser-based remote access, leveraging the Cache Operator to configure port mappings and establish reverse SSH tunnels to the testbed. This ensures a seamless transition from the cloud-native \textit{Federation Plane} to the edge resources.

On the \textit{Testbed Plane}, edge servers provide local storage and computing resources for experiments, for example, baseband signal processing for experiments on the SDR testbed. 
Each server cluster consists of a control plane and multiple worker nodes. The master node coordinates the cluster and hosts several components such as VNC server, SSH tunnel and software configuration, to enable secure remote access from the cloud-based Federation Plane. Each control plane runs an SSM agent to communicate with the Operations Manager, which is deployed in \textit{Federation Plane} with Kubernetes installed. The worker node runs an RKE2 Node agent, which registers the node with the control plane and executes workloads assigned by the Kubernetes. The RKE2-based server node scheduling will be elaborated on in Section~\ref{sec:toolkit}.

\subsection{Toolkit Development} \label{sec:toolkit}
A suite of toolkits has been developed and deployed across the Federation Plane and Testbed Plane, providing a range of services to streamline user experiments. At the core of these services is an automated system that schedules available worker nodes within the edge server cluster and handles automated pod deployment on the selected nodes.

\textbf{RKE2-Powered Worker Node Scheduling.} 
In C-POD, the edge server cluster is orchestrated using RKE2~\cite{rke2docs}, a lightweight single-binary Kubernetes distribution optimized for resource-constrained and unattended edge environments. We chose to use RKE2 because it operates without significant external dependencies and streamlines installation, upgrades, and security hardening through the integration of key components such as containers, Kube-proxy~\cite{kube_proxy}, CoreDNS~\cite{core_dns}, and Container Network Interface (CNI) plugins~\cite{cni_github}. As shown in Fig.~\ref{fig:Podarch}, the RKE2 cluster architecture incorporates the Kube-scheduler and Kubelet components alongside StatefulSet configurations and pods to manage the deployment, scaling and lifecycle of containerized workloads.

The job of the Kube-scheduler is to monitor the cluster for newly created pods without node assignments, evaluate available nodes based on CPU, memory, labels, and other constraints, and rank them according to the scheduling policy. Then, the Kube scheduler interacts with the RKE2 API server to bind each pod to the selected node. Once a pod is assigned, the Kubelet on that node retrieves the corresponding container image from the Image Repository through the built-in RKE2 container runtime called \textit{containerd}. Containerd uses the configured pull secrets and AWS role credentials to access Amazon Elastic Container Registry (ECR), download image layers, mount required volumes, and start the containers.


Then, Kubelet continuously performs readiness and liveness probes, reports pod and node status back to the control plane, and takes corrective actions such as restarting failed pods or managing resources under low-capacity conditions. Once a container is created, the Kubelet configures the probes and resource metrics reporting, after which the container runtime executes a custom startup script. This script establishes a reverse SSH tunnel back to an Elastic Virtual Machine (VM), starts TurboVNC on the node, and launches XFCE4 under a lightweight Xvfb server. On the VM side, a web service based on noVNC listens on the corresponding port, connects through the SSH tunnel to localhost, and renders the XFCE4 desktop of the container in the experimenter browser.

When the system scales or updates a StatefulSet, the control plane repeats the above process: the Kube-scheduler binds new replicas, the Kubelet re-pulls images and re-attaches volumes, and the startup script re-establishes the persistent SSH/VNC bridge, ensuring that each pod delivers an identical GUI-enabled environment.

\textbf{Automated Pod Deployment and Port Management.}
In addition to worker node scheduling, Kubernetes is also leveraged to automate the deployment of a large number of heterogeneous pods. This automation is essential to enhance the resilience and scalability of the C-POD framework. Without it, container instantiation and connection setup would have to be performed manually, which is slow and error-prone, especially when managing a large number of concurrent workloads distributed across multiple testbeds. The challenge becomes even greater when supporting multitasking for multi-user experiments. Preliminary experiments also revealed two critical user experience bottlenecks: (1) SSH tunnels timing out after periods of inactivity, which unintentionally interrupted long-running experiments; and (2) attempts to reconnect to a previous session often triggered the creation of a new pod instead of reengaging with the existing one, unnecessarily consuming resources by provisioning duplicate pods.


To address these challenges, C-POD leverages the Kubernetes API for dynamic provisioning and termination of pods during experiment run-time. When a user initiates an experiment session, the Function Manager module, which works in coordination with the Operations Manager on the Federation Plane (as described in Sec.~\ref{sec:Federation}), launches the necessary functions and procedures to manage users and pods. The C-POD backend then invokes the Kubernetes API to provision a dedicated pod, which will further create an isolated experimental environment tailored to the session’s requirements. 
At the end of the experiment session, the pod and all associated resources are released for reuse. Before calling the API to delete the pod, C-POD first queries Redis to locate the corresponding index entry to ensure proper cleanup. This automated orchestration process enables the timely reclamation of experimental resources while minimizing idle resource utilization. 

\begin{figure}[t]
    \centering
    \includegraphics[width=0.4\textwidth]{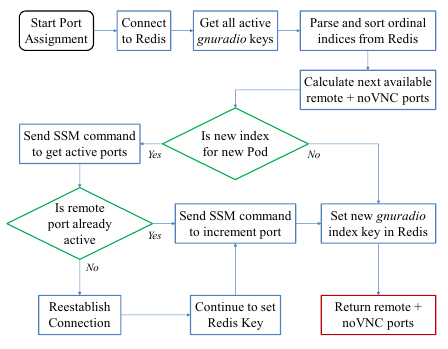}
    \caption{\small Port management workflow for edge server cluster. \vspace{-4mm}
    }
    \label{fig:PortManagement}
\end{figure}

In this process, C-POD manages dynamic port assignments through Redis to avoid conflicts when multiple pods share overlapping remote SSH addresses. As shown in Fig.~\ref{fig:PortManagement}, the workflow is demonstrated using GNU Radio experiments on the SDR testbed. It begins by connecting to Redis and retrieving the set of active GNU Radio keys along with their corresponding indices. These indices are then parsed and sorted in ascending order to identify the next available remote SSH and noVNC ports. If the calculated index corresponds to a new pod, an SSM command is triggered to scale the master cluster while updating the provisioning of the deploying pod. If the remote port is already in use, the system attempts to reclaim it; otherwise, the index is assigned as the next available remote SSH port, and the new pod index is written to Redis. Finally, the assigned remote SSH and noVNC ports are returned to the user.



\textbf{Experiment Session Monitoring.} 
To gain real-time visibility into the health and performance of experiment sessions on edge servers, we deployed a lightweight yet robust monitoring stack based on the open-source Prometheus toolkit and the Grafana visualization platform~\cite{bc2024implementing}. Using Helm, both Prometheus and Grafana were installed onto the Kubernetes cluster. Prometheus continuously collects critical control plane metrics via the \texttt{Kube-apiserver}, including pod memory usage, scheduling delays, and API server request latency. In addition, \texttt{Pushgateway}, \texttt{node-exporter}, and \texttt{Kube-state-metrics} enable comprehensive monitoring at the node and cluster levels. The Prometheus server UI is exposed through the Kubernetes node IP and port, allowing direct access for inspection.

After Prometheus deployment, Grafana was installed and configured to use Prometheus as its data source. The resulting web-based dashboards provide real-time visualizations of experiment session performance, including pod resource utilization, system latency, and workload scaling behavior. These dashboards offer actionable insights into the stability and responsiveness of the C-POD framework, enabling proactive management and rapid troubleshooting during concurrent experiments with multiple users.

\section{Prototyping and Evaluation} \label{sec:evaluation}
We have developed C-POD and deployed it on the AWS cloud platform. Our goals in this section are two-fold: (i) to validate its effectiveness by integrating wireless testbeds at the Testbed Plane and conducting remote experiments on them; and (ii) to evaluate its performance with respect to latency and memory usage during cloud-coordinated pod deployment at the edge testbeds.


\begin{figure}[t]
    \centering
    \includegraphics[
      width=0.35\textwidth,      
      keepaspectratio             
    ]{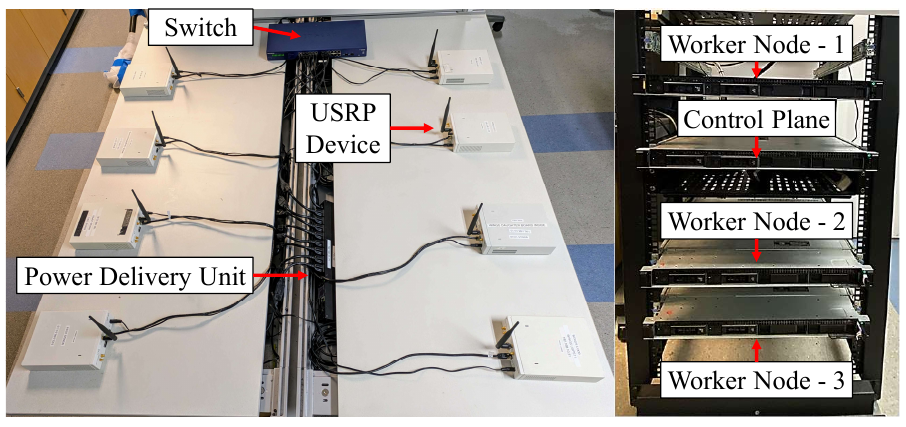}
    \caption{SDR Testbed \vspace{-5mm}}
    \label{fig:Testbed}
\end{figure}

\subsection{Testbed Plane Setup} \label{sec:Testbed Plane}
For demonstration purposes, we integrate two testbeds developed in-house into the C-POD framework: the SDR testbed and the OpenAirInterface (OAI) 5G testbed. As shown in Fig.~\ref{fig:Testbed}, the SDR testbed comprises eight USRP N210 radios, each powered through a Power Distribution Unit (PDU) and interconnected via a Gigabit Ethernet switch. The switch uplinks to a set of edge servers mounted on a rack tray, which run an RKE2‑managed Kubernetes cluster. The cluster consists of a dedicated control-plane node and three worker nodes, each being a Dell workstation equipped with a quad-core Intel Xeon E-2124 CPU at 3.30~GHz and 32~GB of RAM. All edge servers are orchestrated by the C-POD Federation Plane deployed on the AWS cloud. 

\begin{figure}[t]
  \centering
  \begin{tabular}{cc}
    \includegraphics[width=0.2\textwidth]{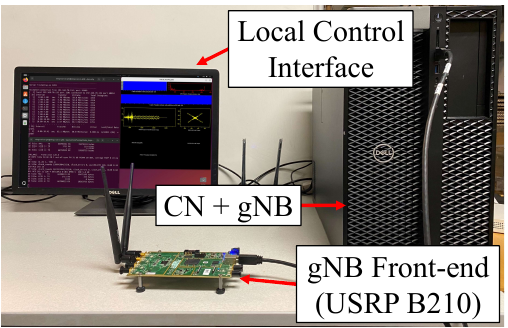} &
    \includegraphics[width=0.2\textwidth]{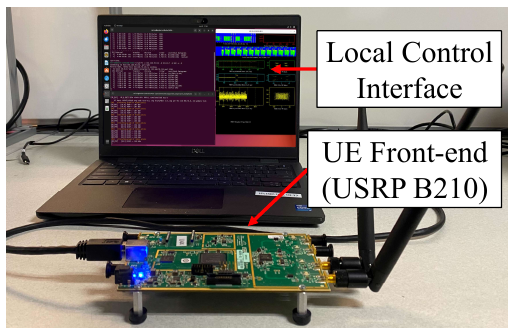} \\
    \small (a)  & \small (b) \\
  \end{tabular}
  \caption{\small OAI 5G Testbed hardware configuration: (a) BS node, comprised of OAI CN and gNB; (b) UE node.
  }
  \label{fig:OAI5G Testbed}
\end{figure}

A snapshot of the OAI 5G \cite{oai5gdocs} testbed is shown in Fig.~\ref{fig:OAI5G Testbed}, illustrating the hardware configurations of Base Station (BS) and User Equipment (UE). The edge server is equipped with a 20-core Intel Xeon Gold 5215 CPU and 64 GB of RAM, which is integrated with the gNB (5G base station) and eNB (LTE base station), radio protocol stack layers such as PHY, MAC, RLC, PDCP, and RRC, and the core network using F1AP or NGAP over the NG interface for 5G. In addition, the 5G core network is compliant with 3GPP standards and is composed of multiple Docker containers including the Network Repository Function (NRF), IP Multimedia Subsystem (IMS), External Data Network (DN), MySQL Database, Unified Data Repository (UDR), Unified Data Management (UDM), Authentication Server Function (AUSF), Access and Mobility Management Function (AMF), Session Management Function (SMF), and User Plane Function (UPF). The UE is equipped with a 12th Gen Intel Core i7-1255U (12 cores) and 32 GB RAM. The current deployment provides one B210 for the gNB and another for the UE. Both BS and UE nodes are orchestrated by C-POD, which enables advanced radio access network experiments.

\begin{figure}[t]
  \centering
  \includegraphics[
    width=0.49\textwidth,
  ]{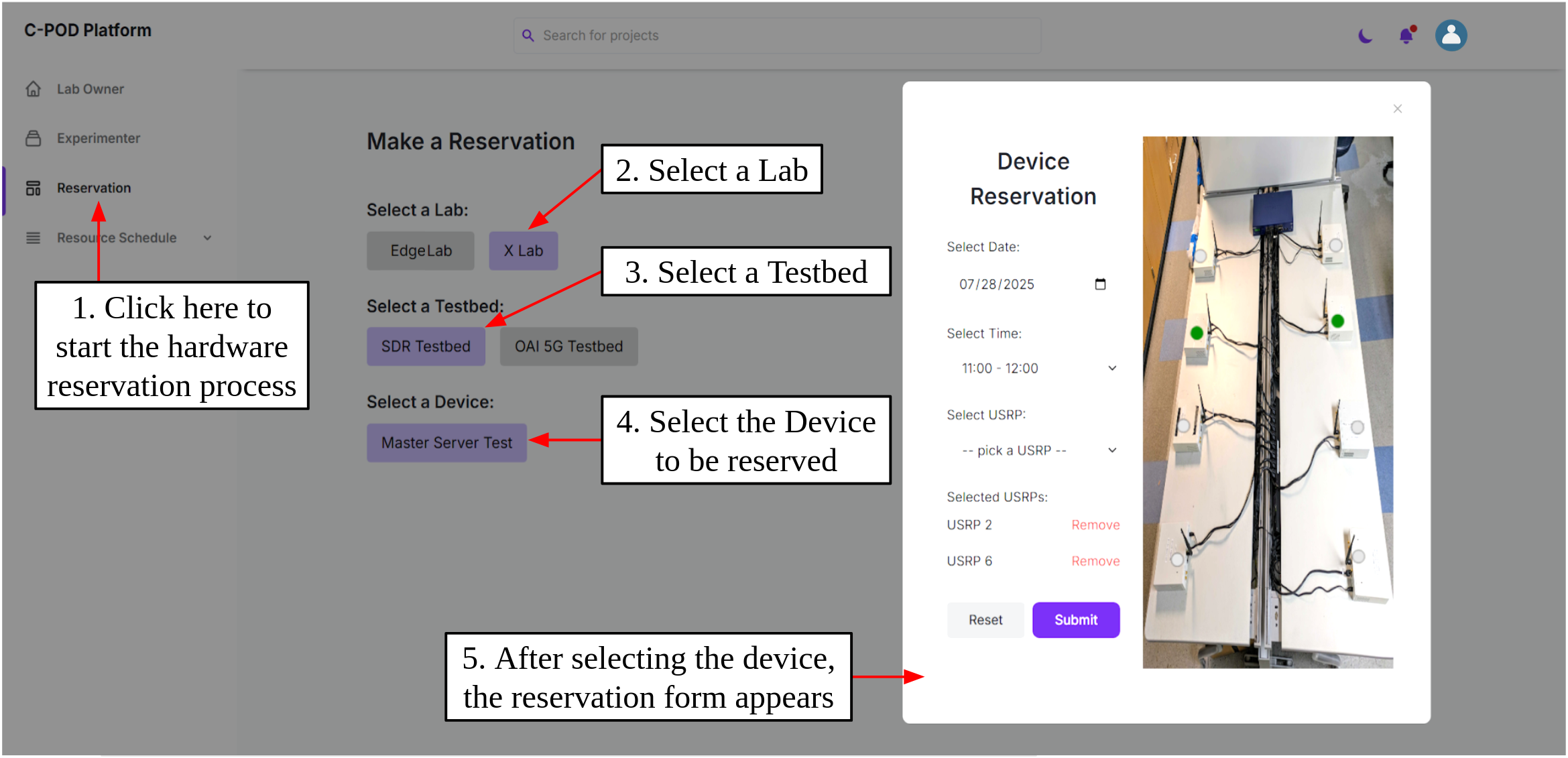}
  \caption{Reservation workflow for a USRP device in the C-POD web portal. \vspace{-4mm}} 
  \label{fig:Reservation}
\end{figure}

\begin{figure*}[t]
    \centering
    \includegraphics[width=0.65\textwidth]{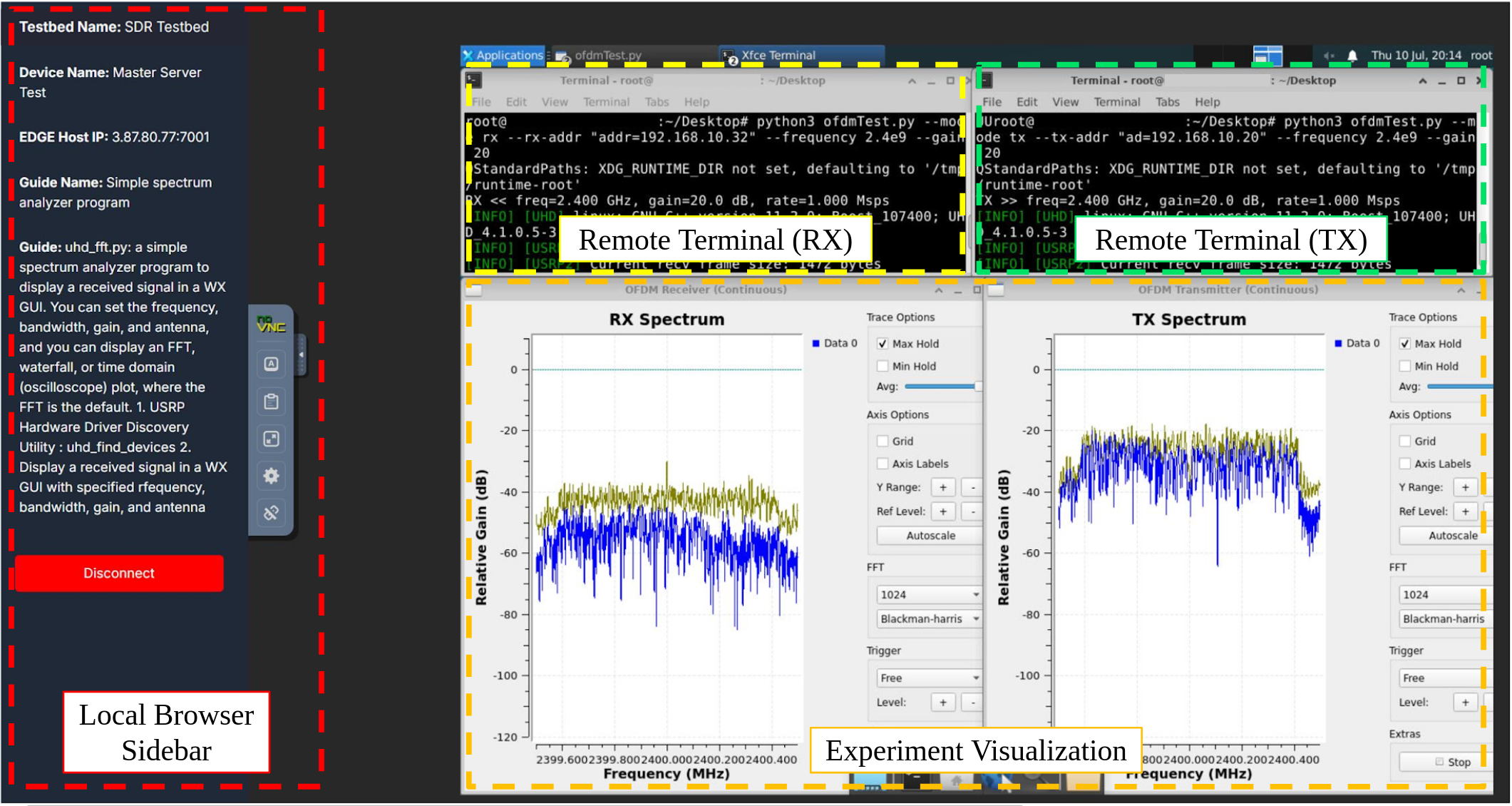}
    \caption{\small Screenshot of remote access to the SDR testbed showing the \textit{RX terminal} (top left), the \textit{TX terminal} (top right; 2.4~GHz, 20~dB gain, 1~Msps), the real-time FFT of the received spectrum (bottom left), and the FFT of the transmitted spectrum (bottom right).} \vspace{-1mm}
    \label{fig:Remote Access SDR}
\end{figure*}

\subsection{Walkthrough Examples} \label{sec:Reservation}

We demonstrate the C-POD workflow through three example experiments: (1) remotely accessing the SDR testbed and reserving SDR resources; (2) running a lightweight experiment that performs an FFT-based spectrum monitoring task over the SDR testbed; and (3) establishing bidirectional communication links using the OAI 5G testbed.

\subsubsection{Resource Reservation} 
The experimenter initiates an experiment by scheduling it through the reservation menu on the C-POD platform’s web portal (\textbf{Step~1}), as illustrated in Fig.~\ref{fig:Reservation}. The workflow proceeds as follows: the experimenter first selects a laboratory from the list of labs connected to C-POD (\textbf{Step~2}), then chooses the testbed available within that lab (\textbf{Step~3}). In this example, the SDR testbed in the X Lab is selected. Next, the experimenter reserves the computing and SDR resources associated with the chosen testbed (\textbf{Step~4}).  

As shown on the right-hand side of Fig.~\ref{fig:Reservation}, selecting an edge server opens a reservation form displaying all the SDR resources linked to that server, accompanied by a calendar interface. To further streamline the process, a graphical visualization of the lab’s testbed layout has been integrated into the form. This intuitive interface allows the experimenter to specify the desired date, time, and device for the experiment, with the selected USRP clearly highlighted in the visualized experiment scenario. For example, in Fig.~\ref{fig:Reservation}, the two selected USRPs are indicated with green dots.

\subsubsection{Remote Access to SDR Testbed and Experiment} 
During the reserved time slot, the user navigates to the experiment section in the sidebar, where a \textit{Connect} button becomes available. Clicking \textit{Connect} opens a modal dialog listing all active session instances and offering the option \textit{Open New Instance}. Selecting this option will launch the remote experiment interface, which presents a virtual desktop environment equipped with an embedded terminal for running experiments.
Then, the experimenter can initiate their experiments there. 

In this example, the transmit USRP continuously generates and transmits Quadrature Phase Shift Keying (QPSK) mapped Orthogonal Frequency Division Multiplexing (OFDM) signals at the selected frequency. The receive USRP captures the OTA I/Q samples and displays a live Fast Fourier Transform (FFT) trace. These FFT traces enable the user to monitor spectral occupancy, evaluate bandwidth utilization, and characterize the wireless channel in real time. It is important to note that this lightweight experiment is intended solely to demonstrate remote access to the edge testbed through the C-POD platform. Experimenters can also upload and execute their own custom-designed scripts for more advanced experiments using the file upload functionality provided by C-POD.


Fig.~\ref{fig:Remote Access SDR} presents a screenshot of an ongoing experiment. On the left side of the interface, the local browser displays the testbed name, device name, host IP, and a concise experiment guide. This panel serves as a convenient reference during the session, allowing the user to track the experimental context and follow instructions without referring to external documentation. In the main area, the virtual desktop shows the execution of the transmit and receive scripts within the embedded terminal, while the plots below display the measured RX and TX spectra, providing visual confirmation of real-time transmission and reception over the selected frequency band. Upon completion of the experiment, the user can terminate the session by clicking the \textit{Disconnect} button in the sidebar, ensuring that resources are properly released and made available for subsequent experiments.

\subsubsection{Link Establishment on Remote OAI 5G Testbed} 
C-POD can be used to conduct experiments involving multiple remote terminals on the target testbed. As first, as shown in Fig.~\ref{fig:Reservation}, the experimenter navigates to the \textit{Reservation} section in the sidebar and selects the corresponding laboratory and testbed (OAI 5G in this case). The experimenter can then establish access to two edge servers and launch a remote experiment terminal on each, enabling parallel control and monitoring the transmission and reception behaviors of the gNB and UE. 
The screenshot is omitted due to space constraints.



\begin{figure}[t]
    \centering
    \includegraphics[width=0.35\textwidth]{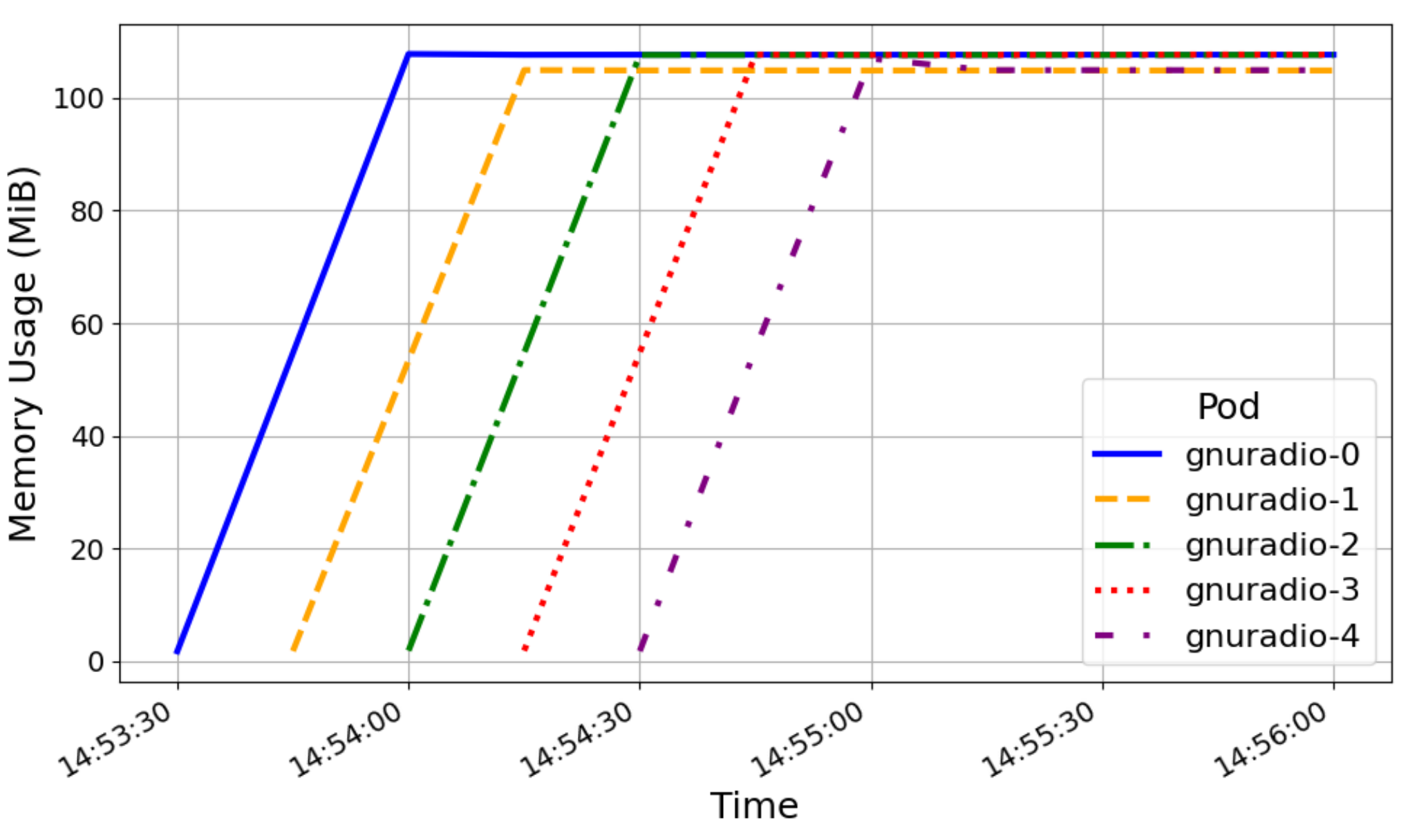}
    \caption{\small Memory usage of pods over time, illustrating a consistent ramp-up phase followed by stable allocation.
    \vspace{-5mm}} 
    \label{fig:Memory consumption by pods}
\end{figure}


Within the terminals, both the gNB and UE softmodems have been executed to establish a complete standalone 5G NR link operating on FR1 band~78 with 106~PRBs and 30\,kHz subcarrier spacing. The gNB continuously transmits synchronization and control channels, manages real-time PHY/MAC scheduling, processes uplink PUSCH transmissions from the UE, and demodulates PUCCH feedback, while simultaneously reporting real-time link metrics including RSRP, SNR, BLER, MCS selections, and HARQ statistics. On the UE side, the device locks onto the downlink broadcast synchronization, compensates for frequency offsets, performs an RRC attach with its IMSI, transmits uplink data bursts, and records its own RSRP, HARQ feedback, and throughput measurements. On both ends, I/Q samples of the RF signals are streamed into XForms-based scopes, providing time- and frequency-domain resource grids, channel impulse responses, log-likelihood ratios, constellation diagrams, and instantaneous bit-rate visualizations.

\begin{figure}[t]
    \centering
    \includegraphics[width=0.32\textwidth]{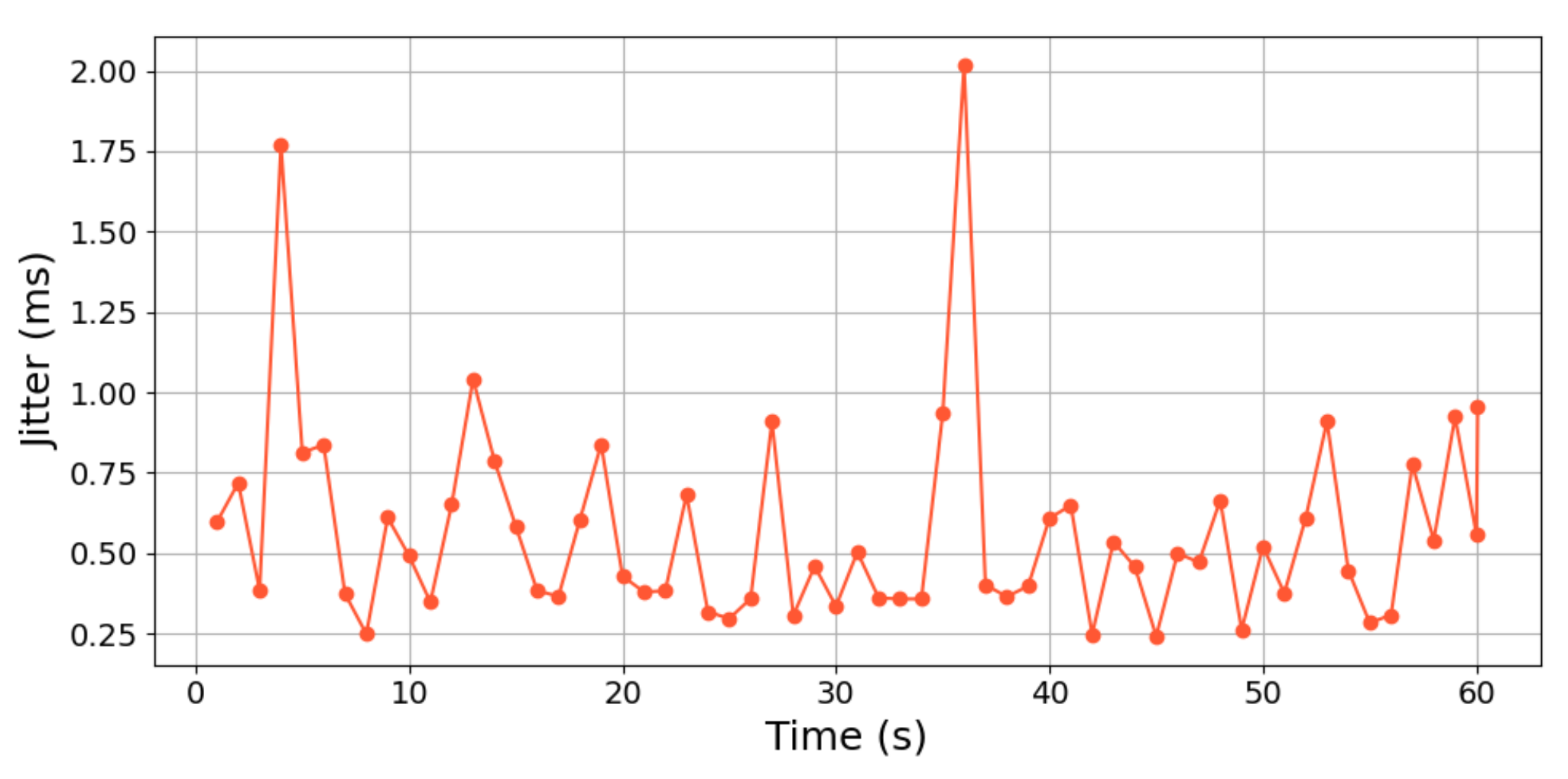}
    \caption{\small Latency between Federation Plane and deployed pods. \vspace{-4mm}} 
    \label{fig:Latency Measurement}
\end{figure}

\subsection{Scalability of Pod Deployment} \label{sec:scalability}

In this experiment, we evaluate the effectiveness of C-POD in supporting multi-pod deployments for ongoing experiments and assess its scalability in terms of edge memory utilization and lifecycle management. The evaluation is performed in a Kubernetes cluster deployed on the edge servers of the SDR testbed, which consists of three worker nodes, each equipped with a quad-core Intel Xeon E-2124 @ 3.30\,GHz and 32\,GB of RAM, and a single control-plane node. Pods are incrementally deployed to characterize how the system allocates resources and maintains performance under increasing workloads.

We incrementally deploy five pods (\texttt{gnuradio-0} through \texttt{gnuradio-4}), each running on the worker nodes and configured to request 2\,GB memory and 500\,mCPU. As shown in Fig.~\ref{fig:Memory consumption by pods}, each pod exhibits a consistent and predictable memory utilization pattern. Specifically, after instantiation, the
memory usage increases smoothly during initialization until it reaches the configured 2\,GB request, after which it stabilizes at a steady level.
No significant spikes, drops, or signs of contention appear as additional pods are launched.

\subsection{Cloud-Edge Latency Measurement} \label{sec:latency}
To evaluate the stability and predictability of latency between the cloud and the edge node, we measure the per-interval jitter of UDP traffic flowing from the federation plane to the edge server. By analyzing jitter, defined as the variability in packet interarrival times, we assess how reliably the system supports real-time interaction with the wireless testbeds.



Specifically,
we generate a continuous 10 Mbps UDP stream from the AWS cloud to the edge pod using the open source network performance tool iperf3. Packet arrival timestamps are captured at the edge with the command line packet analyzer tcpdump to display the network traffic, and the traces are analyzed offline to compute inter-arrival jitter.
As shown in Fig.~\ref{fig:Latency Measurement}, the observed jitter remains consistently low, typically ranging from 0.3~ms to 0.8~ms, with only two spikes exceeding 1.75~ms around the 4-5~s and 36-37~s marks. The overall average jitter was approximately 0.6~ms, with the 95th percentile staying below 1.0~ms. These results indicate that packet arrival times varied slightly, suggesting that 
C-POD is able to process incoming UDP flows with very little buffering.
Even under sustained 10\,Mbps UDP load, the network path maintains sub-millisecond variability for the vast majority of the test duration. This low and predictable jitter profile demonstrates that the edge-in-pod deployment provides reliable, low-latency connectivity to the Federation Plane. 
\vspace{-1mm}
\section{Conclusions} \label{sec:Conclusion}

In this work, we have presented C-POD, a cloud-powered framework that automates the deployment and management of edge pods to enable scalable and accessible OTA experimentation on heterogeneous wireless testbeds. Using public cloud resources and distributed edge infrastructure, C-POD lowers the barrier to OTA experimentation and enables seamless testbed sharing without requiring extensive local setup. A prototype deployed on AWS and integrated with RF sensing and 5G testbeds demonstrates consistent resource utilization, uniform pod startup behavior, and reliable horizontal scaling. 

\bibliographystyle{ieeetr}
\bibliography{Annoy, GuanPub, unionlabs}
\end{document}